\documentclass[preprint,1p,times,longtitle]{elsarticle}

\usepackage[utf8x]{inputenc}
\usepackage{cite}
\usepackage{url}
\usepackage{amsmath,amssymb,amsfonts}
\usepackage{algorithmic}
\usepackage{graphicx}
\usepackage{textcomp}
\usepackage{xcolor}
\def\BibTeX{{\rm B\kern-.05em{\sc i\kern-.025em b}\kern-.08em
    T\kern-.1667em\lower.7ex\hbox{E}\kern-.125emX}}

\usepackage[nolist]{acronym}
\usepackage{todonotes}
\usepackage{dirtytalk}
\usepackage{multirow}
\usepackage{amsthm}
\usepackage{booktabs}
\usepackage{listings}
\usepackage{hyperref}

\usepackage[T1]{fontenc}
\usepackage[scaled=0.9]{beramono}
\usepackage{microtype}

\newcommand\Small{\fontsize{8}{10}\selectfont}
\newcommand*\LSTfont{%
  \Small\ttfamily\SetTracking{encoding=*}{-50}\lsstyle}

\presetkeys{todonotes}{fancyline,color=green}{}

\newcommand{\oslc}[1]{\emph{oslc:#1}}
\newcommand{\trs}[1]{\emph{trs:#1}}
\newcommand{\auto}[1]{\emph{oslc\_auto:#1}}

\newcommand{\actions}[1]{\emph{oslc\_eca:#1}}
\newcommand{\events}[1]{\emph{oslc\_eca:#1}}

\newcommand{\kafka}[1]{\emph{kafka:#1}}
\newcommand{\eca}[1]{\emph{oslc\_eca:#1}}

\newpageafter{author}

\begin{document}

\begin{frontmatter}

\title{Extending the OSLC standard for ECA-based automation in DevOps environments}

\author[gsi]{Guillermo García-Grao\corref{cor1}}
\ead{g.ggrao@upm.es}

\author[gsi]{Álvaro Carrera}
\ead{a.carrera@upm.es}

\cortext[cor1]{Corresponding author}
\address[gsi]{Universidad Politécnica de Madrid, Avenida Complutense 30, Ciudad Universitaria, 28040, Madrid}


\begin{abstract}

The DevOps paradigm is taking over software development systems, helping businesses increase efficiency, accelerate production, and adapt quickly to market changes.
However, adopting these principles can be challenging.
Practitioners often face an important issue known as vendor lock-in caused by the cost of tool replacement.
In addition, automating the processes that involve these tools also requires investment.
These issues could be addressed by standardizing service interfaces to facilitate their integration.
Linked Data is an attractive choice for implementing such a standard without sacrificing versatility.
An exciting and promising proposal in this direction is the \ac{OSLC} standard specification.
Its purpose is to build an environment where services can interoperate using standard Linked Data models.
However, the current specification version still lacks standard definitions for concepts that are critical to automating the execution of actions in fast-changing environments.
Therefore, this paper proposes a new specification to extend \ac*{OSLC}, based on the \ac*{ECA} model, for event-based interoperable automation, especially for DevOps environments, which are our motivational scenario.
A simple DevOps architecture is built as a prototype to validate the proposed model.
Using that architecture, the proposed model is validated in a real-world workflow to prove its contribution to the OSLC standard and the DevOps field.

\end{abstract}

\begin{keyword}
OSLC \sep DevOps \sep Linked Data \sep semantic \sep ontology \sep rule-based automation
\end{keyword}

\end{frontmatter}

\begin{acronym}
 
 \acro{OWL}[OWL]{Web Ontology Language} 
 \acro{RDF}[RDF]{Resource Description Framework}
 \acro{RDFS}[RDFS]{Resource Description Framework Schema}
 \acro{OSLC}[OSLC]{Open Services for Lifecycle Collaboration}
 \acro{EWE}[EWE]{Evented WEb}
 \acro{TAS}[TAS]{Task Automation Server}
 \acro{ECA}[ECA]{Event-Condition-Action}
 \acro{AaaS}[AaaS]{Automation as a Service}
 \acro{W3C}[W3C]{World Wide Web Consortium}
 \acro{MQTT}[MQTT]{Message Queuing Telemetry Transport}
 \acro{TRS}[TRS]{Tracked Resource Set}
 \acro{EYE}[EYE]{Euler Yet another proof Engine}
 \acro{CAD}[CAD]{Computer-Aided Design}
 \acro{BPMN}[BPMN]{Business Process Modeling Notation}
 \acro{CRUD}[CRUD]{Create, Read, Update and Delete}
 \acro{FOL}[FOL]{First Order Logic}
 \acro{CI}[CI]{Continuous Integration}
 \acro{DML}[DML]{Distributed Messaging Layer}

\end{acronym}

\section*{Acknowledgment}

This research work is supported by the Ministry of Science and Innovation (Spain) through the SmartDevOps project (RTC2019-007326-7). 

\section{Introduction}\label{sec:introduction}

In recent years, a set of tools, practices, and philosophies in the software development industry have become increasingly popular. Their goal is to improve efficiency in the application lifecycle process, resulting in faster production and delivery. This paradigm, known as DevOps~\citep{jabbariWhatDevOpsSystematic2016}, requires an infrastructure capable of rapidly scaling and adapting to changes as the market moves constantly. It also needs to accelerate the delivery of services and applications by automating tasks in phases such as testing and deployment, a practice known as Continuous Delivery~\citep{chenContinuousDeliveryOvercoming2017}.

Therefore, the line between development and operations teams is becoming thinner as many companies begin to interact with their infrastructure programmatically, using the APIs that their tools and services provide, seeking a more versatile and scalable way of managing them~\citep{artacDevOpsIntroducingInfrastructureasCode2017}. Integrating these tools provides flexibility and allows practicioners to automate processes, resulting in faster and more efficient production. Such automation enhancements are becoming more attractive to the point where tools that work themselves as automation providers for other services are being implemented.

The tools that provide automation capabilities are known as \ac{TAS}~\citep{coronadoTaskAutomationServices2016}, and most of them are based on the \ac{ECA} model. This model consists of three phases: an event triggers a rule, then a condition is evaluated, and finally, an action is executed accordingly. The concept implemented by \ac{TAS}s is known as \ac{AaaS}. Examples of such technologies, which focus on giving users the ability to define automation for their applications and processes, are IFTTT~\citep{iftttIFTTT} or Zapier~\citep{ZapierAutomationThat}. In the context of software development, tools such as StackStorm~\citep{StackStorm} apply the same ideas to automate production and delivery tasks. These tools differ from other workflow execution frameworks like Apache Airflow or GitHub Actions in approaching rule definition. In these frameworks, users are provided with tools to build their workflows over the specific platform they are using. For example, in GitHub Actions, even though it can be integrated with external tools, users are meant to build their workflows based on the activity of a GitHub repository. Instead, a \ac{TAS} focuses on integration with any services that are able to generate events and execute actions on demand. The users can use the \ac{TAS} to build rules based on these external resources.

Nevertheless, there are still two critical challenges associated with the adoption of DevOps in general and \ac{AaaS} in particular: the potential elevated cost of integrating new services~\citep{eversEvaluatingCloudAutomation2015} and the lack of flexibility when automating processes~\citep{kargerSemanticWebEnd2014}. They are analyzed in more depth in later sections. A potential solution to the integration problem found in the literature~\citep{wettingerStandardsBasedDevOpsAutomation2014} would be standardizing the interfaces between the different tools, making the migration process fast and seamless. However, this is not a simple task. Every tool has its particular model, architecture, and requisites, so it would be difficult for any possible standard to adjust appropriately to all of them. Furthermore, even if such a standard was implemented, forcing vendors to adapt their services to a rigid model could create more difficulties. They would lose their freedom to experiment with new features and functionalities.

Standardizing how the involved services are modeled could also be beneficial in terms of possible flexibility improvements for automation solutions. Again, the challenge is to design a sufficiently flexible model so that new services can be directly integrated without requiring them to adjust to a restrictive protocol.

In this context of interoperability between services, solutions based on semantic web technologies have been considered before~\citep{mcilraithSemanticWebServices2001}. Following the principles of Linked Data, a service can get heterogeneous data from different sources and still process it, a feature that could simplify integrations between services.

There are already projects attempting to standardize the interfaces between lifecycle tools using Linked Data, the most relevant for our purposes being \ac{OSLC}~\citep{amsdenOSLCCoreVersion2021}. Because its models are not fixed and can be expanded, various tools can define their specific concepts and vocabularies. At the same time, other services can still understand and work with their resources. For example, suppose that a vendor implements a new feature that cannot be represented with its current classes. In that case, they can create an ontology and expand their resources with new definitions. Other services can discover this feature by following the links in the resource definitions, avoiding many of the usual integration issues.


Similarly, semantic web technologies have also been applied in the context of \ac{TAS}. \ac{OSLC} has a specific domain for automation tools. However, it has some limitation when it comes to modeling \ac{ECA}-based environments. Later sections explore this topic in more detail. The main issue is that \ac{OSLC} lacks a proper semantic model for events and actions.

The contribution of this paper is to define a new semantic model for \ac{OSLC} that extends the Automation domain to support events and actions. The model provides standard interaction with \acp{TAS} based on \ac*{ECA} rules. More specifically, it has been designed to apply to real-world DevOps scenarios. To validate this last claim, an architecture prototype is presented based on the proposed semantic model to face some usual challenges DevOps practitioners face. Finally, a worked example has been conducted to evaluate this environment.

The remainder of this paper is structured as follows. Section~\ref{sec:related_work} presents related works that have laid the basis for the presented proposal. Section~\ref{sec:background} explains the concepts defined by \ac*{OSLC} and their limitation when modeling an \ac*{ECA} environment. In Section~\ref{sec:semantic_model} the proposed semantic model for extending \ac{OSLC} is proposed. Section~\ref{sec:evaluation} analyzes the main challenges in DevOps adoption and defines an architecture to face them based on the \ac*{OSLC} extension. It also presents a worked example to evaluate the proposal's validity in a real-world scenario. Finally, the conclusions and possible future work are described in Section~\ref{sec:conclusions_and_future_work}.

\section{Related work}\label{sec:related_work}

This section presents research work in the field of knowledge related to this paper. It starts by reviewing ontologies proposed in the literature for DevOps environments. Special attention is payed to those that focus on the challanges explained in Section~\ref{sec:introduction}. \ac{OSLC} stands out for its approach to interoperability. Then, a study is conveyed to find architectures where the \ac{OSLC} standard has been used, emphasising those where automation is the main focus.



As mentioned before, an interesting approach to the problem of integrating tools in a DevOps environment is to use semantic technologies~\citep{mcilraithSemanticWebServices2001}. These techniques can provide a flexible representation of resource data that different programs and devices can seamlessly understand. However, defining an ontology model for DevOps is not a simple task, as it is a concept related to different domains. Moreover, the very definition of DevOps is still imprecise and a subject of many debates, so the attempts to create such a vocabulary are often either excessively abstract and general or too specific for a particular tool. A definition based on the concepts found in the literature that has been proposed is the following: \say{DevOps is a development methodology aimed at bridging the gap between Development (Dev) and Operations, emphasizing communication and collaboration, continuous integration, quality assurance, and delivery with automated deployment utilizing a set of development practices}~\citep{jabbariWhatDevOpsSystematic2016}.

There are case studies where the use of Linked Data in a system of systems has been used for a road construction project~\citep{axelssonExperiencesUsingLinked2019}. A core ontology was defined and made extensible using word models. In another work~\citep{johngHarmoniaContinuousService2019}, a DevOps ontology extracted from Jenkins logs was used. Concepts where defined like Deployment Task or Infrastructure Change, but it is warned that this is just a reference ontology and might not be generalizable to other scenarios.

Another example of semantic technologies used with DevOps~\citep{mccarthyComposableDevOpsAutomated2015} establishes a common language for operations and development teams to use and defines some ontologies for the maturity of DevOps. More examples include a framework based on a DevOps ontology~\citep{guerreroSystematicMappingStudy2020} built on top of the PrMO (Ontology of Process-reference Models)~\citep{pardo-calvacheReferenceOntologyHarmonizing2014} and the SMO (Software Measurement Ontology)~\citep{barcellosWellFoundedSoftwareMeasurement2010}. Finally, the PaaSport semantic model~\citep{bassiliadesPaaSportSemanticModel2018}, an extension of the DUL (DOLCE+DnS Ultralite) ontology~\citep{gangemiUnderstandingSemanticWeb2003}, has been reviewed. However, it is more focused on compatibility between cloud computing platform providers.

Different Linked Data approaches to the tool interoperability problem in software development projects have been analyzed before~\citet{rodriguezFormalOntologiesData2019}. One of them was \ac{OSLC}. The \ac{OSLC} project~\citep{amsdenOSLCCoreVersion2021} tries to define a standardized interface for software development tools that applies the semantic technologies provided by \ac{W3C}. \ac{OSLC} is a set of open specifications for tool integration using the Linked Data initiative. The project consists of a Core specification that every \ac{OSLC} compliant tool must follow and a series of domain-specific protocols that are more suited for the particular needs of some groups of tools.


Based on \ac{OSLC}, a modeling tool for the specification of Linked Data resources focused on the development toolchain has been proposed~\citep{el-khouryModellingSupportLinked2016}. However, the authors also state that their modeling tool should be extended to cover other aspects of DevOps, like requirements analysis and automated testing.

The combination of standardization and flexibility found in \ac{OSLC} made it suited for the purpose of this paper. To learn more about previous applications of it, different architectures based on \ac{OSLC} from the literature are explore. The following case studies help illustrate how versatile \ac{OSLC} is and how different domains could benefit from it being more widely adopted.



Beyond the software engineering field, architectures integrating IoT devices using \ac{OSLC} have been proposed~\citep{chenOpenSourceLifecycle2019}. Because of the heterogeneity of these IoT devices and the data they generate, getting them to interoperate becomes a problem. Using Linked Data and a standard protocol is helpful in this situation. 

Another \ac{OSLC}-based architecture is used to integrate testing and requirements tools~\citep{nardoneOSLCbasedEnvironmentSystemlevel2020} in the context of the European Rail Traffic Management System/European Train Control System. Furthermore, the implications that \ac{OSLC} architectures have on implementation in the context of space systems have also been discussed~\citep{hoppeRequirementsSharedData2015}. Finally, \ac{OSLC} has been used to connect \ac{CAD} design tools and product visualization~\citep{ebelingOSLCBasedApproach2017}.

Modeling also presents interoperability challenges because system models are usually composed of submodels from different tools. For example, combinations of \ac{OSLC} and OpenModelica for model management and traceability~\citep{mengistTraceabilitySupportOpenModelica2017} or uses of the \ac{OSLC} Knowledge Management specification to allow lifecycle artifacts to be reused~\citep{alvarez-rodriguezElevatingMeaningData2019}.
In addition, the PROMIS project~\citep{aoyamaPROMISManagementPlatform2013} uses \ac{OSLC} to allow for an interorganizational software development model with a Software Supply Network (SSN).


A previous work that implemented an interoperability platform based on \ac*{OSLC} is the Crystal project~\citep{leitner_lessons_2016}. It was tested in a larger and more complex case study than the one presented in this paper. However, although \ac{OSLC} is used as the foundation for the project, their models go far beyond the proposed \ac*{OSLC} standards. This paper aims to define a model that could serve as a potential new specification in \ac{OSLC}. Also, the semantic models they proposed are not explicitly designed for \ac*{ECA} automation, which is the primary purpose of the present work.

There is a previous article that presents an architecture that includes something similar to event-based automation with \ac*{OSLC}~\citep{berezovskyiImprovingLifecycleQuery2018}. It uses it in the context of an integrated toolchain for IoT and uses the \ac{TRS}~\citep{crossleyOSLCTrackedResource2021}, an \ac*{OSLC}-based protocol, to automate different tasks.
They affirm that this approach would also be very beneficial in Big Data environments because \say{using microservices and Linked Data to integrate systems within the toolchain eliminates the need for multiple transformations}, achieving better velocity and variety than previous techniques. They propose a case study of a robotic warehouse optimized for logistics automation. It uses a few tools from the change management and requirements management domains, each with its own \ac{OSLC} adapter. To communicate changes in these tools, instead of using HTTP, they resort to the \ac{MQTT} protocol~\citep{banksMQTTVersion2019} to have everything synchronized.

The proposal is exciting because it integrates various tools and devices using semantic technologies and creates a centralized communication network between these tools, which offers the possibility to automate their behavior.
They even make an effort to deliver messages asynchronously, essential for faster and more efficient communication.
However, the semantic models it uses have not been designed explicitly for \ac*{ECA}, as \ac*{TRS} is a change tracking protocol, but it was not created with events in particular.
The following section explores \ac*{OSLC} in more detail to provide a better context and to highlight the limitations found when it comes to automation based on event-based models.

\section{Background}\label{sec:background}

This section explains how the \ac*{OSLC} Core Specifiaction works in more detail. It also presents \ac{OSLC} Automation and \ac{TRS}, domains that are relevant in the motivational DevOps scenarios. First, Section~\ref{sec:service_modelling} introduces the Core specifaction of \ac*{OSLC} which is the foundation for every service modeling. Then, Section~\ref{sec:auto_modelling} showcases the Automation domain which would be the best suited to model a \ac*{TAS} in the current state of the project. Next, Section~\ref{sec:logging_modelling} presents the \ac*{TRS} protocol, useful for logging and change tracking. Finally, Section~\ref{sec:oslc_limitations} discusses some of the limitations found in these vocabularies when trying to use them to model an \ac*{ECA} environment.





\subsection{OSLC Core overview}\label{sec:service_modelling}

\ac{OSLC} is a family of open specifications for integrating tools. Its purpose is to integrate product/application lifecycle services regardless of their internal model. This goal makes the standard suitable for modeling the services in a DevOps infrastructure. In \ac{OSLC}, different topics such as change management, test management, or requirements management define specifications for their respective resources. They are known as domains. These domains are built on top of the \ac*{OSLC} Core Specification~\citep{amsdenOSLCCoreVersion2021} to ensure their compatibility.

In \ac{OSLC}, data is represented using \ac{RDF} and retrieved via HTTP\@. Therefore, each service must provide a catalog of its Service Providers, represented by the \oslc{ServiceProviderCatalog} and \oslc{ServiceProvider} classes. The \ac{OSLC} primer defines Service Providers as \say{organizing concepts that partition the overall space of artifacts in the tool into smaller containers. Examples of common partitioning concepts offered by tools include \say{projects}, \say{modules}, \say{user databases}, and so on}~\citep{speicherOSLCPrimer2019}. Hence, the meaning of Service Provider changes for each service, but the main idea is that it classifies the resources in that service.

\begin{figure}[!ht]
    \centering
    \includegraphics[width=0.9\textwidth]{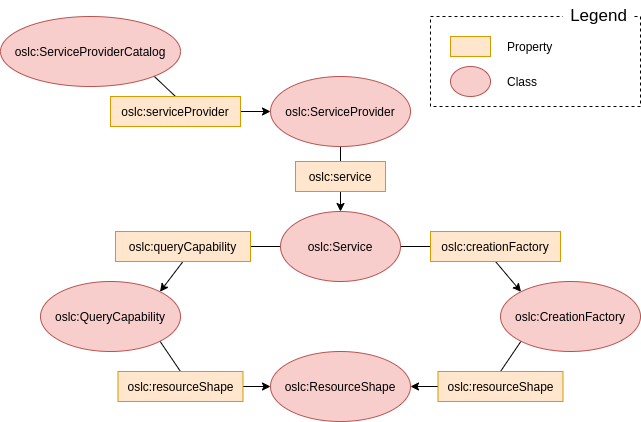}
    \caption{\ac{OSLC} Core overview.}\label{fig:oslc_core_overview}
\end{figure}

An \ac*{OSLC} API offers clients the possibility of discovering all its resources and capabilities based on its root resource, the \oslc{ServiceProviderCatalog}. This resource allows clients to obtain its list of Service Providers responsible for providing the URLs needed to send creation and query requests to the server. These endpoints can be discovered through the \oslc{CreationFactory} and \oslc{QueryCapabilities} services. They may also provide an \oslc
{ResourceShape} that describes the properties a resource from that service is expected to have. An overview of the \ac{OSLC} Core specification from \citet{amsdenOSLCCoreVersion2021} can be found in Figure~\ref{fig:oslc_core_overview}.



\subsection{OSLC Automation}\label{sec:auto_modelling}

\begin{figure}
    \centering
    \includegraphics[width=0.8\textwidth]{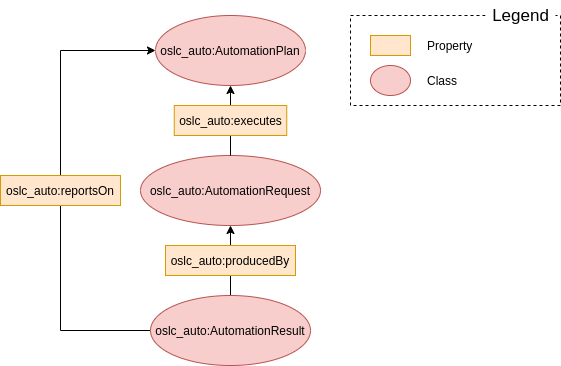}
    \caption{\ac{OSLC} Automation diagram.}\label{fig:oslc_automation_overview}
\end{figure}

Regarding the process automation field, \ac{OSLC} defines a specific domain called \ac{OSLC} Automation~\citep{amsdenOSLCAutomationVersion2021}. This domain aims to model an interface so that an automation provider can interact with other \ac{OSLC} services. It defines three different resources a Service Provider exposes following the Core standard. The first of these resources is the \auto{AutomationPlan} which defines a unit of automation available for execution. Next, the \auto{AutomationRequest} resource provides the information required to execute an Automation Plan. Finally, the standard defines an \auto{AutomationResult} resource to track an Automation Request status and contributions. Figure~\ref{fig:oslc_automation_overview} displays a diagram of these resources and their relationships.

\subsection{The TRS protocol}\label{sec:logging_modelling}

As briefly mentioned in Section~\ref{sec:related_work}, the \ac{TRS} protocol~\citep{crossleyOSLCTrackedResource2021} is specified as part of the \ac{OSLC} project. It was conceived to track changes in the resource set of an \ac{OSLC} service and expose them as \ac{RDF} HTTP resources themselves. These exposed changes have been used before to trigger automated actions~\citep{berezovskyiImprovingLifecycleQuery2018}.

\ac{TRS} defines a list of \trs{ChangeLog} resources. Each of these Change Logs contains numbered entries representing every creation, modification, or deletion suffered by any resource in the set. Because this list can be accessed via HTTP, a client can periodically check what changes have occurred in the resource set to monitor the behavior of the service. The protocol also offers the \ac{TRS} Patch mechanism to allow a \trs{ChangeEvent} to carry more detailed information about the modifications suffered by the resources.

\subsection{OSLC limitations for ECA-based automation}\label{sec:oslc_limitations}

After reviewing the current state of the \ac*{OSLC} project, the goal is to introduce it in a DevOps environment. More specifically, an environment where a \ac*{TAS} provides automation for other services using a system of \ac*{ECA} rules. To achieve this, \ac*{OSLC} needs a standardize way for services to provide event detection and actions execution capabilities.

As previously mentioned, there is a specific \ac*{OSLC} domain for Automation. However, key concepts are still missing for this model to support event detection. For example, to run an \auto{AutomationPlan}, \ac*{OSLC} Automation requires a consumer to create an \auto{AutomationRequest} by making a POST request to the Automation server. The ideal scenario would be to have the \ac*{OSLC} Automation service listening for events generated in another service by itself. This idea is the starting point for defining events in the proposed extension model.

On the other hand, \ac{OSLC} resources already support basic CRUD actions (create, read, update and delete). It was already mentioned how Service Providers enable resource creation and querying. The resources in \ac{OSLC} can also be updated or deleted using the corresponding HTTP methods (PUT and DELETE, respectively). The \ac{OSLC} community has shown efforts to support the execution of more concrete actions~\citep{painOSLCActions2020} over resources. The proposal allows a service to perform actions on other service resources without knowing their type and properties. The document is a potential candidate for a future specification. It even suggests ideas of introducing actions in the \ac{OSLC} Core Specification. This draft serves as foundation for the action model proposed in this paper, extending to improve some key aspects still missing.
Next section presents the extension to \ac*{OSLC} for event and action support. It details all the properties that have been defined and explians how it is suposed to be used for interoperability.

\section{OSLC Extension for ECA-based Automation}\label{sec:semantic_model}

This section showcases the main contribution of the paper which is the semantic model that extends \ac*{OSLC} to support \ac*{ECA}-based automation~\citep{OSLCExtensionECAbased}. Section~\ref{sec:event_modeling} details how events are proposed to work. Section~\ref{sec:action_modeling} presents an overview of the new actions vocabulary. Finally, Section~\ref{sec:rules} explains how automation rules are defined and provides a whole overview of the propoed \ac*{ECA} model. For the state of clarity, every new concept proposed in the paper uses the \textit{oscl\_eca} prefix.

\subsection{Event modeling}\label{sec:event_modeling}

\begin{figure}
    \centering
    \includegraphics[width=\textwidth]{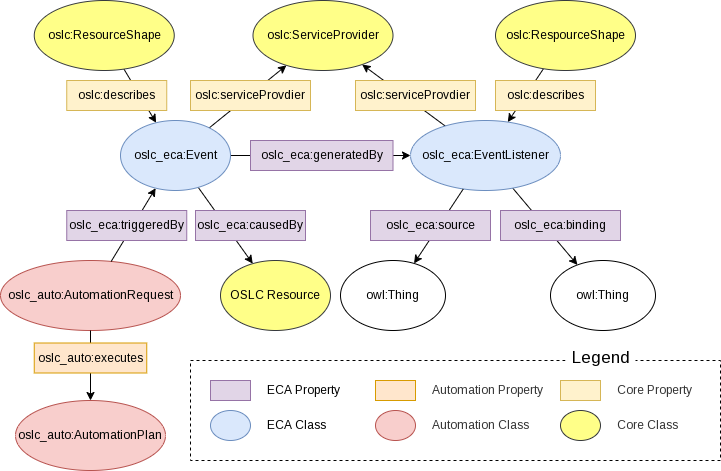}
    \caption{Diagram of classes and properties for Events - OSLC ECA proposal.}\label{fig:oslc_events}
\end{figure}

This section details the psoposed model for integration events in \ac{OSLC}. A diagram is shown in Figure~\ref{fig:oslc_events} with these newly defined resources.

The philosophy followed by the \ac*{OSLC} standard is not to make assumptions about its tools and services. Therefore, the focus has to be on the concepts enabling interoperability. In the context of events, a concept needs to be defined so consumers can discover how to listen to a service generating them. For this purpose, the \events{EventListener} resource is defined.

An \events{EventListener} resource represents a backgorund process insida the service that listens to a source of events. Consumers can create an \events{EventListener} so it notifies them when an event occur. The source of these events could be many things depending on the service and is signaled by the \events{source} property. Other properties, required or optional, can be described by an \oslc{ResourceShape}, as in any other \ac*{OSLC} resource.

However, unlike other \ac*{OSLC} resources, interacting with it using CRUD actions is not enough. Although a consumer could periodically poll the service for new events, the optimal result would be for the service to asynchronously notify consumers. There are multiple options to do this, among others: using webhooks, providing a URL to the service to make POST request with the new events; with a publish-subscriber protocol, like MQTT which has been used before in a similar context~\citep{berezovskyiImprovingLifecycleQuery2018}; or using some messaging service that can deliver the events to their intended receivers.

There are too many possiblities to standardize all of them. To solve this issue, a concept known as interacion patterns can be borrow from the \ac*{OSLC} Actions proposal~\citep{painOSLCActions2020}. The idea is to determine which protocol to use depending on the type of resource pointed by a \events{binding} property. The simplest example is with a webhook. When the \events{binding} property is supposed to point at a \textit{http:Request}, then it means the service uses webhooks. A consumer could then create an \events{EventListener} providing a description of the \textit{http:Request} that wants to receive. This leaves the door open to new interaction patterns to join the standard in the future.

The events themselves are represented by an \ac{OSLC} resource named \events{Event}. Whenever an \events{EventListener} detects an event triggered by its sources, it creates and exposes an \events{Event}. Depending on the interaction pattern used by the service, it may also send the \events{Event} asynchronously (inside POST request, for example). Besides the \ac*{OSLC} stabdard properties and typical metadata annotations (like a title or a timestamp), an \events{Event} resource can have other service specific properties. An \oslc{ResourceShape} should describe all of the possible properties the event can contain. An \events{Event} also has the \events{generatedBy} and \events{causedBy} properties pointing at the \events{EventListener} and \ac*{OSLC} resource that produced it, respectively.

There is a relation between these events and the \ac*{OSLC} Automation domain. For example, an \events{Event} could be used to trigger \auto{AutomationPlans}. Hence, there should be a relationship between \events{Events} and \auto{AutomationRequests}. An \auto{AutomationRequest} could be started when the automation service receives an \events{Event}, so a new \events{triggeredBy} property is proposed for addition.

Before going into the next secton, another interaction pattern more suitable for DevOps scenarios needs to be explored. The webhook example served to illustrate how a simple interaction pattern could be defined but is not ideal for environments where scalability is critical. Using a message broker to deliver the events to consumers is a more appropriate solution. Tools like Apache Kafka~\citep{ApacheKafka} provide this kind of service and allow for more scalable architectures. For this reason, this interaction pattern is used in the architecture and worked example presented in Section~\ref{sec:evaluation} for the evaluation of the model.

Such an interaction pattern requires the \events{binding} to point at a resource representing the broker, with properties signaling a URL for providers to POST messages and a topic for conumers to subscribe. The \textit{kafka} prefix is used to model this service. The \kafka{Broker} resource is proposed to model the broker, along with the the \kafka{providerURL} and \kafka{topic} properties.


    

\subsection{Action modeling}\label{sec:action_modeling}

\begin{figure}
    \centering
    \includegraphics[width=\textwidth]{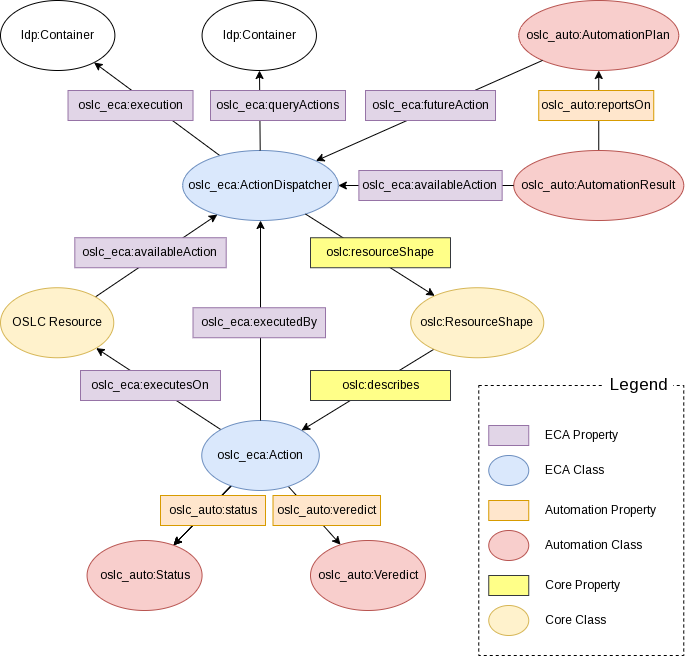}
    \caption{Diagram of classes and properties for Actions - OSLC ECA proposal.}\label{fig:oslc_actions}
\end{figure}

This section presents the proposed model for actions in \ac{OSLC}.
Figure~\ref{fig:oslc_actions} shows a diagram representing its classes and properties, relations with other \ac{OSLC} concepts.

The starting point for the action model is the \ac{OSLC} Actions proposal, previously mentioned in Section~\ref{sec:background}. According to it, resources may have an \actions{action} property advertising possible actions that can be performed on them.
This property leads to \actions{Action} resources that describe the action and provide instructions to execute it and determine its result.
The information is provided through action bindings which can support multiple interaction patterns (i.e., HTTP request or UI dialog).
This is the same method mentioned for the events model proposed in the previous section. Making each resource expose its own available actions facilitates interoperability.

However, in many cases, resources from the same \oslc{ServiceProvider} and same domain have the same available actions.
It would be interesting to provide a way of discovering available actions for a whole set of resources in a Service Provider, instead of having to query specific resources to find them.
Also, the binding and interaction pattern mechanisms seem overly complicated in this case, as \ac*{OSLC} already has the concept of \oslc{ResourceShape} to indicate to consumers how to work with resources like \actions{Actions}.

A different approach is proposed to solve these issues. The \actions{ActionDispatcher} resource is defined to represent the capability of the service to execute a specific action. Inspired by the \oslc{CreationFactory} and \oslc{QueryCapabilities} concepts, an \actions{ActionDispatcher} provides a URL where consumers can send post requests to execute the action it dispatches. The \actions{execution} data property signals this URL. Actions sent for execution can be queried using another URL provided by the \actions{ActionDispatcher} through the \actions{queryActions} predicate. These URLs are represented by \textit{ldp:Container} objects, like in the \ac*{OSLC} Core specification.

\actions{ActionDispatchers} also provide an \oslc{ResourceShape} to let consumers know the necessary properties to execute its \actions{Actions}. Individual resources can still indicate which actions are available for execution by pointing with an \actions{availableAction} to the corresponding \actions{ActionDispatcher}. Furthermore, an executed \actions{Action} points at its target resource with the \actions{executedOn} property. Finally, the \actions{Actions} contain a \actions{status} and \actions{verdict} properties indicating the execution result, similar to an \auto{AutomationRequest}.

The standard establishes a relation between the Automation and Actions specifications. Specific \actions{Actions} can become available (or be automatically executed) when an \auto{AutomationPlan} successfully finishes running. The \actions{futureAction} property was defined to identify them. In this extended model, \actions{futureAction} points to an \actions{ActionDispatcher} instead. The corresponding \auto{AutomationResult} indicates that the \actions{Action} is available for immediate execution.





\subsection{Rule modeling}\label{sec:rules}

\begin{figure}
    \centering
    \includegraphics[width=\textwidth]{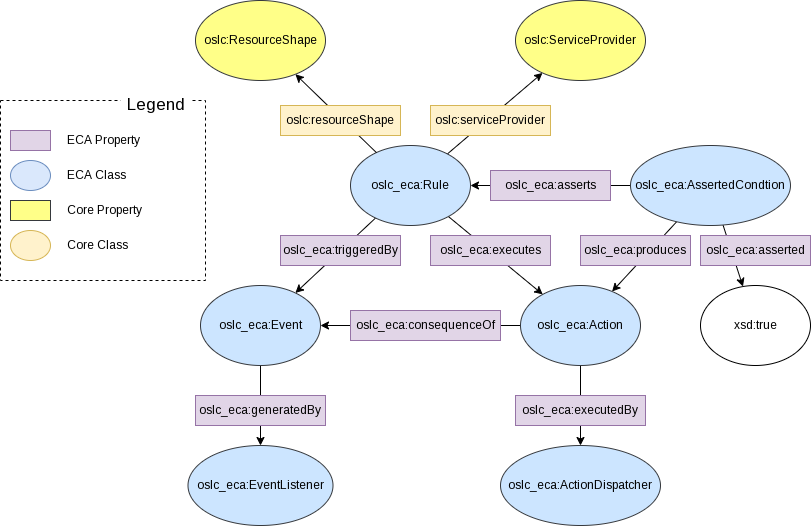}
    \caption{Diagram of classes and properties for Rules - OSLC ECA proposal.}\label{fig:oslc_rules}
\end{figure}

Besides events and actions, it is necessary to have a model for automation rules to fully represent an \ac*{ECA} environment.
These rules represent the conditions that need to be met for an event to trigger an automated action.
Figure~\ref{fig:oslc_rules} shows the whole model proposed in this work.

An \eca{Rule} represents an automation rule.
It can point to an \events{Event} with an \eca{triggeredBy} property and to an \actions{Action} with an \eca{executedBy} predicate.
The \eca{consequenceOf} property is also defined for actions to point at events that caused them to be executed.

The \eca{AssertedCondtion} resource is defined to provide a means of discovering if some evaluation of a rule was successful.
When the rule is evaluated, an \eca{AssertedCondtion} is created pointing at the specific \actions{Action} generated with the \eca{produces} property.
The \eca{asserts} predicate links it to the \eca{Rule}.
A boolean linked by the \eca{asserted} property indicates whether the evaluation was successful. 

Because \ac*{OSLC} is focus on interoperability, the proposal is not concerned with the way automation rules work internally.
Depending on the service managing the automation, rules can be evaluated in many ways.
They can even use one of the frameworks that exist for semantic rules, like SPIN~\citep{knublauch2011spin} or Notation 3 Logic~\citep{berners2005notation}.
This would mean further interoperability, as they are also Linked Data standards.
Therefore, the proposed rule modeling aims to establish a framework able to cope any rules based on the \ac{ECA} model without specifying the method applied during the inference process.

\section{Validation}\label{sec:evaluation}

This section discusses the validity of the results obtained by the proposed semantic model.
The need for these extension to the \ac*{OSLC} standard was justified in Section~\ref{sec:introduction} with the benefits it provides to the DevOps field.
Section~\ref{sec:problem_overview} explores the literature to find notable issues faced by DevOps practitioners.
Based on these challenges, requirements are defined for a DevOps architecture.
Meeting these requirements means the architecture improves practitioners' experience regarding the issues found in the literature.

To prove that \ac*{OSLC} and, more specifically, the proposed semantic model help in fulfilling the requirements, a DevOps architecture is presented in Section~\ref{sec:architecture}.
The concepts defined in Section~\ref{sec:semantic_model} are used to build an automation environment based on the \ac{ECA} model.
To ensure the architecture actually follows the DevOps philosophy, a set of characteristic for such architectures are gathered from the literature.
The most relevant are included in the design process.

Finally, to validate the proposed semantic model, the architecture enabled by it is put to the test with a worked example conducted in Section~\ref{sec:case_study}.
It presents a real-world scenario where two popular commercial services are integrated into the \ac*{ECA} environment.
It proves that the model and the architecture it made possible meet the established requirements.

\subsection{Problem overview}\label{sec:problem_overview}

This section focuses on establishing a set of requirements to face the challenges identified in the DevOps domain. The goal is to show how the proposed model can be used to build an architecture that meets the requirements and, therefore, helps improving DevOps practices. The list challenges has been gathered from a systematic review of the literature~\citep{bolscherDesigningSoftwareArchitecture2019}. The ones addressed by this proposal have been selected and are shown in Table~\ref{tab:issues}. They also put together a list of characteristics desirable for DevOps environments that will be useful for the design of the architecture.

\begin{table}[!ht]
    \centering
    \begin{tabular}{ccc}
        \toprule
        \textbf{ID} & \textbf{Issue} & \textbf{Description} \\ \midrule
        IS1 & Methods and tools &
        \begin{tabular}[c]{@{}c@{}}
          Complex tooling and lack \\
          of proper support
        \end{tabular} \\ \midrule

        IS2 &
        \begin{tabular}[c]{@{}c@{}}
          Ever-changing operational \\
          environments and tools
        \end{tabular} &
        \begin{tabular}[c]{@{}c@{}}
          Deploying in heterogeneous \\
          operations environments
        \end{tabular} \\ \midrule

        IS3 & Logging & Bug traceability with many services \\ \midrule

        IS4 & Monitoring & Complex and sophisticated \\ \midrule

        IS5 & Scaling &
        \begin{tabular}[c]{@{}c@{}}
          Hard to implement DevOps if the \\
          architecture does not scale
        \end{tabular} \\ \bottomrule
    \end{tabular} 
\caption{List of issues/challenges for DevOps environments~\citep{bolscherDesigningSoftwareArchitecture2019}.}\label{tab:issues}
\end{table}

Section~\ref{sec:introduction} mentions some issues found when adopting DevOps practices. First, the problem known as vendor lock-in appears when software companies become dependent on the tools they are using, not being able to substitute them when they need to~\citep{opara-martinsCriticalAnalysisVendor2016}. Companies seeking to adopt DevOps practices like \ac{CI} could face this challenge due to the complexity of the required tools and the effort needed to integrate them into their workflows~\citep{stahlCindersContinuousIntegration2017} (issue IS1). Even those companies that have already successfully transitioned to DevOps could suffer from this issue, as the environments and tools are changing fast and constantly~\citep{shahinIntersectionContinuousDeployment2016} (issue IS2). This issue results in a lack of flexibility that is incompatible with the DevOps idea of quickly adapting to changes in the market.

Regarding \ac{AaaS}, designing flexible solutions can be a problematic task~\citep{coronadoTaskAutomationServices2016}. For most \ac{TAS} platforms, workflows are defined based on available features. Regardless of how extensive they make this set of features, there can be situations where users need some specific functionality that the vendor does not offer. Users might, for example, want to integrate a recently released service into their infrastructure, but it may not be yet supported by the \ac{TAS} they are currently using~\citep{coronadoTaskAutomationServices2016}.

Even when users implement such integrations themselves, they might want to switch to another \ac{TAS} later. However, because they have all of their automated tasks built on top of the \ac{TAS} they are currently using, they might find it too expensive to make the migration, getting again to a position of vendor lock-in. These issues could lead to companies not adapting and evolving fast enough, so their infrastructure quickly becomes obsolete.

Another issue that arises when adopting DevOps is managing logging and monitoring (issues IS3 and IS4). In a microservices architecture, traceability becomes increasingly hard to handle at scale. Failing to address traceability adequately could delay finding bugs which means less efficient \ac{CI}.

While facing the previous challenges, other DevOps concepts should not be overlooked. Flexible implementations and a short time to market are key DevOps aspects, but the scalability of the infrastructure (issue IS5) should not be sacrificed to achieve them.

These concepts will be the foundation for the defined requirements. Other works presenting DevOps architectures make quantitative analyses to evaluate the performance gained by their proposed approach~\citep{vergoriDevOpsPerformanceEngineering2017}. However, in this case, it is hard to find values that can be measured to make such assessments. Therefore, the proposed model is evaluated by defining a set of requirements, presenting an architecture based on such a model, and implementing a test case to verify that they are correctly satisfied.


To tackle the service integration issue (composed of IS1 and IS2), requirement \textbf{RQ1} is defined. Because changes in the development and business model can happen over time, tools will need to be added, updated, or removed from the infrastructure. The integration between these tools should be as independent of their internal structure as possible to maximize versatility. Tools need to provide a means to interact with them in a standardized way, either directly or through an adapter. For the architecture proposed in this paper, that standard will be \ac{OSLC}. This requirement can be validated by successfully executing a complete workflow that detects an event in one service, triggers a rule in a \ac{TAS}, and executes an action in a different service. Such a workflow should be implemented without changing the system interfaces. They have to work regardless of the internal structure of the integrated service.


Requirement \textbf{RQ2} focuses on the flexibility challenges faced by \ac{AaaS} platforms. To avoid the vendor lock-in problem, it should be possible to interact with the \ac*{TAS} using a standardized protocol. In this case, that standard is \ac*{OSLC}. In particular, it uses the concepts from the \ac*{OSLC} Automation domain.
\textbf{RQ3} addresses the problem of logging and monitoring (IS3 and IS4) the components of the architecture. Once again, using a standard would simplify this task. In additon, traceability becomes easier if every service follows the same model for its logs. The \ac*{TRS} standard fits because of its compatibility with \ac*{OSLC}.


Growth in the complexity of the infrastructure is supported as the number of tools integrated increases. \textbf{RQ4} guarantees that the architecture can scale and adapt to such changes without being restructured (IS5). To satisfy this requirement, the system should rely on technologies able to scale horizontally. In other words, as long as the different modules can be containerized and provide APIs to access them, the infrastructure can be considered scalable.



The following section presents an architecture based on the proposed semantic model to fulfill these requirements, summarised in Table~\ref{table:reqs}.

\begin{table}
    \centering
    \begin{tabular}{lp{0.5\textwidth}l} 
        \toprule
        \textbf{RQ ID} & \textbf{Description} &\textbf{Related Issues}\\ 
        \hline
        RQ1         & The infrastructure shall seamlessly integrate any tool, as long as it follows specific standards. & IS1, IS2 \\ 
        \hline
        RQ2         & The system shall use a standardized interface to communicate with automation services. & IS1, IS2 \\ 
        \hline
        RQ3         & The system shall use a standard for logging and monitoring. & IS3, IS4 \\ 
        \hline
        RQ4         & The infrastructure shall be scalable and support larger scenarios. & IS5 \\
        \bottomrule
    \end{tabular}
    \caption{Summary of defined requirements and their relation to the issues found in the literature for DevOps.}\label{table:reqs}
\end{table}

\subsection{Prototype Architecture}\label{sec:architecture}

This section presents an architecture that uses the \ac*{OSLC} extension model as a foundation. 


This section presents an architecture founded on the \ac*{OSLC} extension model. The decisions made in the architecture design are aimed to meet the requirements established in Section~\ref{sec:problem_overview}. Because it is meant to be applicable in DevOps scenarios, it also includes the essential characteristics deemed beneficial for such an architecture in the literature~\citep{bolscherDesigningSoftwareArchitecture2019}. The most relevant of these characteristics have been selected to guide its design. The criteria used are based on the number of references and the relation to the requirements of this proposal. These chosen characteristics are listed in Table~\ref{tab:characteristics}. Some characteristics not selected could be included in future work.

\begin{table}[!ht]
    \centering
    \begin{tabular}{cccl}
        \toprule
        \textbf{ID} & \textbf{Characteristic} & \textbf{Related RQs} & \textbf{Description} \\ \midrule

        CH1 & Deployability & RQ1 and RQ4 &
        \begin{tabular}[l]{@{}l@{}}
            The architecture deploying downtime is \\
            minimized and moving between different \\
            environments is fast.
        \end{tabular} \\ \midrule

        CH2 & Modularity & RQ1 and RQ4 &
        \begin{tabular}[l]{@{}l@{}}
            Services in the architecture have minimized \\
            dependencies and changes are isolated.
        \end{tabular} \\ \midrule

        CH3 & Loosely coupled & RQ1 and RQ4 &
        \begin{tabular}[l]{@{}l@{}}
            Decreasing application and inter-team \\
            dependencies.
        \end{tabular} \\ \midrule

        CH4 & Agility/Modifiability & RQ1 &
        \begin{tabular}[l]{@{}l@{}}
            Architectural changes can be performed fast.
        \end{tabular} \\ \midrule

        CH5 & Automation & RQ2 &
        \begin{tabular}[l]{@{}l@{}}
            Processes like testing and deployment \\
            are automated as much as possible.
        \end{tabular} \\ \midrule

        CH6 & Monitoring & RQ3 &
        \begin{tabular}[l]{@{}l@{}}
            Services can be monitored.
        \end{tabular} \\ \midrule

        CH7 & Logging & RQ3 &
        \begin{tabular}[l]{@{}l@{}}
            Services produce and store meaningful logs.
        \end{tabular} \\ \bottomrule
    \end{tabular} 
\caption{List of beneficial characteristics for DevOps architectures by \citet{bolscherDesigningSoftwareArchitecture2019}.}\label{tab:characteristics}
\end{table}

\begin{figure}[!ht]
    \centering
    \includegraphics[width=0.9\textwidth]{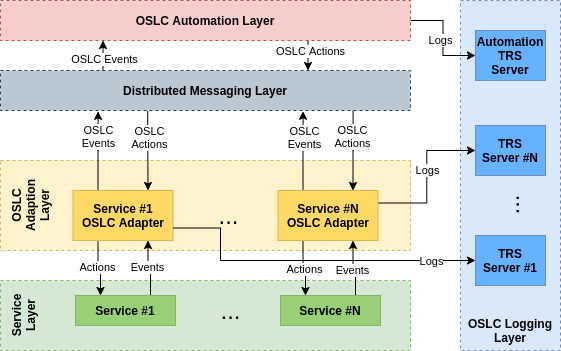}
    \caption{Diagram of the prototype architecture for validation.}\label{fig:basic_architecture}
\end{figure}

The architecture comprises five layers, as shown in Figure~\ref{fig:basic_architecture}. These five layers are the Service Layer, the \ac*{OSLC} Adaption Layer, the Distributed Messaging Layer, The \ac{OSLC} Automation Layer, and the \ac{OSLC} Logging Layer. In the following paragraphs, the inclusion of each layer is justified by the defined requirements and the selected characteristics.

The most frequently required characteristic in a DevOps environment is deployability (CH1). A deployable architecture needs to be easily moved between environments. This is often achieved by dividing it into smaller components that can be deployed independently (microservices). Modularity (CH2) and loosely coupled (CH3) are two more characteristics this pattern meets. In the proposed architecture, these components form the \textbf{Service Layer}.

To meet requirement RQ1, services in the architecture need to be able to communicate via \ac*{OSLC}. This would also achieve modifiability (CH4), as services would be easier to substitute if their interfaces are standardized. Some of these services might be natively compliant with \ac*{OSLC}, but those not following the standard require adapters. These adapters are independent component, so the previously mentioned characteristics are not sacrificed. They form the \textbf{\ac*{OSLC} Adaptation Layer}.

Automation (CH5) is another crucial characteristic for DevOps practitioners. It is fulfilled, along with requirement RQ2, by the \textbf{\ac{OSLC} Automation Layer}. To keep the architecture modular, automation is separated into its own service. It is compliant with \ac*{OSLC} and communicates with the other services through \ac*{OSLC} adapters. However, having the \ac{OSLC} Automation Layer directly communicate via HTTP requests with the adapters would cause issues as the system scales and complexity grows.

To avoid conflicts with requirement RQ4, a \textbf{\ac{DML}} is included. This service is responsible for delivering the messages generated in the services to the \ac{OSLC} Automation Layer and vice versa. Any service in the architecture can push messages to the \ac*{DML} that will be received by those subscribed to the appropriate topic.

Finally, to meet requirement RQ3, every \ac*{OSLC}-compliant service implements the \ac*{TRS} specification. \ac*{TRS} provides tracking capabilities for changes in a service's resource set. This capability can be used for monitoring (CH6) and logging (CH7). The \ac*{TRS} functionality is separated from the service into its own component to keep a modular architecture. The collection of \ac*{TRS} services form the \textbf{\ac{OSLC} Logging Layer}.

\subsection{Worked Example of an ECA-based automation in the motivational scenario}\label{sec:case_study}

This section presents a worked example with a practical architecture implementation involving two real-world services.
Its goal is to demonstrate that the proposed model enabled an architecture that can meet the requirements defined in Section~\ref{sec:problem_overview}.
The issues and requirements defined in that section are addressed throughout this worked example.
The chosen tools are Bugzilla~\citep{Bugzilla}, a bug tracking web service, and GitHub \citep{GitHub}, a git-based version control tool.
Both tools are well known and widely used in the software development industry.
Also, \ac*{OSLC} already provides a Change Management domain to model bug tracking tools.
This presents an opportunity to show how the model integrates with other parts of \ac*{OSLC}.
All the code is available in a public repository\footnote{Worked example source code: \url{https://github.com/gsi-upm/oslc-eca_environment}}.

A practical example is introduced where this integration could be helpful. A company that has already adopted some DevOps practices. The development team uses GitHub repositories to manage the code of the applications they are working on. They are also using GitHub issues for change management. GitHub offers integration with many tools, and the development team uses some of them for CI/CD\@. For example, they have an automated workflow that deploys a new version of an application when a pull request to the main branch is accepted. This workflow is only executed if there are no open issues with a \say{critical} label.

Working on the same project, another team is in charge of testing the application. This team wants to use Bugzilla instead of GitHub issues to keep track of the bugs and errors they find. A reason they want this could be, for example, because Bugzilla is open source and can be hosted by them. The development team needs to be aware of this new tool and integrate it into their automated workflows, so the application is not deployed while there are unresolved bugs (issue \textbf{IS1}). To save the work of manually doing this, they could use some already developed solutions to integrate GitHub and Bugzilla, such as GitZilla~\citep{geraGitZilla2021}. However, this approach makes them more dependent on these tools. As established by issue \textbf{IS2}, the tools and environments in DevOps change all the time. The testing team might want to move to a different bug tracking tool in the future. To avoid these issues and meet the requirement \textbf{RQ1}, the company decides to use \ac*{OSLC} standardized interfaces to integrate both tools.

Adapters must be built because none of these services are natively integrated with \ac{OSLC} or \ac{TRS}.
This tool is based on the \ac*{ECA} model: it receives events from other services, evaluates internal rules, and sends back actions for execution.
An adapter has already been implemented for Bugzilla as part of a tutorial on \ac{OSLC} made by some Core committee members\footnote{OSLC Bugzilla tutorial: \url{https://oslc.github.io/developing-oslc-applications/integrating_products_with_oslc/running_the_examples}}. Bugzilla classifies its \say{bugs} into \say{products}. Hence, the adapter chooses these elements as its \ac{OSLC} resources and \oslc{ServiceProviders}. It operates in the Change Management domain. Whenever a bug is created or updated, it is reflected in the \ac{TRS} server provided by the adapter.

GitHub has no \ac{OSLC} capabilities implemented but provides a REST API with all the necessary features to implement an \ac{OSLC} server. For this worked example, the resources of interest on GitHub are its \say{issues}, as they are closely related to bugs in Bugzilla. The concept of \say{repository} is chosen as \oslc{ServiceProvider} and Change Management as the domain. GitHub also offers a webhook-based notification system whenever an event occurs. The \ac{TRS} server can be updated without periodically polling the API to check for changes with this feature.

The development team uses a \ac*{TAS} to implement its automation rules keeping its architecture modular.
Because the rest of the tools they are using are integrated with \ac*{OSLC}, they implement an \ac*{OSLC} interface to interact with the \ac*{TAS} in a standardized way. These decisions are consistent with the requirement \textbf{RQ2}. Three workflows are defined to synchronize the services using these events and actions. Table~\ref{tab:table_workflows} illustrates how these workflows and their respective rules operate.

\begin{table}[!ht]
    \centering
    \begin{tabular}{cccc} 
        \toprule
        \multicolumn{2}{l}{\multirow{2}{*}{}} & \multicolumn{2}{c}{Workflows}                                                               \\ 
        \cmidrule(lr){3-4}
        \multicolumn{2}{l}{}                   & Bugzilla to GitHub                         & GitHub to Bugzilla                            \\ 
        \midrule
        \multirow{3}{*}{Rules} & Create        & \(Bug\:created \rightarrow Create\:issue\) & \(Issue\:created \rightarrow Create\:bug\)    \\ 
        \cmidrule(lr){2-4}
                               & Update        & \(Bug\:updated \rightarrow Update\:issue\) & \(Issue\:updated \rightarrow Update\:bug\)  \\
        \cmidrule(lr){2-4}
                               & Resolve        & \(Bug\:resolved \rightarrow Close\:issue\) & \(Issue\:closed \rightarrow Resolve\:bug\)  \\
        \bottomrule
    \end{tabular}
    \caption{Workflows and rules defined for the worked example.}\label{tab:table_workflows}
\end{table}

To support standardized logging and monitoring of their services (\textbf{RQ3}), they use \ac*{TRS} servers to expose changes in every \ac*{OSLC} resource. In addition, they use Apache Kafka to send messages between servicest to maintain scalability as their automation environment grows (\textbf{RQ4}).

\begin{figure}[!ht]
    \centering
    \includegraphics[width=\textwidth]{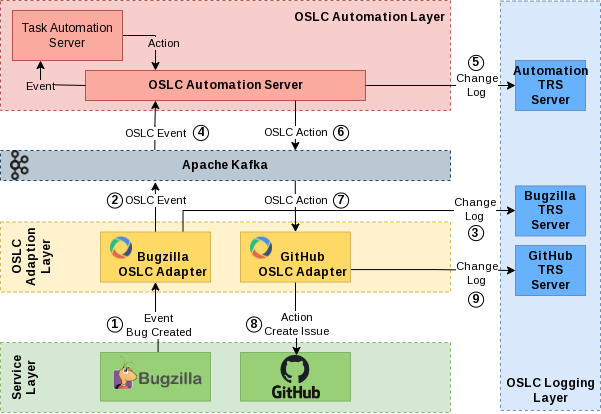}
    \caption{Workflow detailed in the worked example, where the creation of a bug in Bugzilla triggers the generation of an issue in GitHub.}\label{fig:case_study}
\end{figure}

Figure~\ref{fig:case_study} shows the complete implementation of the architecture for the exposed worked example. The diagram shows the workflow where a bug is created using Bugzilla, and an issue is automatically generated on GitHub following the \ac*{OSLC} \ac*{ECA} proposed extension.

Following the diagram steps (Figure~\ref{fig:case_study}), the workflow starts with a bug creation in Bugzilla (Step 1).
To describe resources for bug tracking like this, \ac*{OSLC} provides the Change Management domain~\citep{amsdenOSLCChangeManagement2020}.
The \ac*{OSLC} Bugzilla adapter uses this domain's \textit{oslc\_cm:ChangeRequest} class to model bugs.
Listing~\ref{listing:bug} shows the example resource generated from the created bug in Notation 3 language.

\begin{lstlisting}[caption={\ac{RDF} representing the bug created in Bugzilla.}, label={listing:bug}]
@prefix bgz: <http://www.bugzilla.org/rdf#> .
@prefix oslc: <http://open-services.net/ns/core#> .
@prefix dcterms: <http://purl.org/dc/terms/> .
@prefix oslc_cm: <http://open-services.net/ns/cm#> .
@prefix xsd: <http://www.w3.org/2001/XMLSchema#> .

<http://bugzilla.example.com/ServiceProvider/1/bugs/1> a oslc_cm:ChangeRequest ;
    oslc_cm:severity "enhancement"^^xsd:string ;
    oslc_cm:status "CONFIRMED"^^xsd:string ;
    oslc:serviceProvider <http://bugzilla.example.com/ServiceProvider/1> ;
    dcterms:contributor "admin"^^xsd:string ;
    dcterms:created "2021-12-10T07:07:33+00:00"^^xsd:dateTime ;
    dcterms:identifier "1"^^xsd:integer ;
    dcterms:modified "2021-12-10T07:07:33+00:00"^^xsd:dateTime ;
    dcterms:title "Testing bug to issue automation"^^xsd:string ;
    bgz:component "TestComponent"^^xsd:string ;
    bgz:operatingSystem "Linux"^^xsd:string ;
    bgz:platform "PC"^^xsd:string ;
    bgz:priority "---"^^xsd:string ;
    bgz:version "unspecified"^^xsd:string .    
\end{lstlisting}

The creation is notified to the Event Listener of its \ac*{OSLC} adapter, which sends an \ac*{OSLC} modeled event containing all the bug's properties to the Apache Kafka server (Step 2).
Listing~\ref{listing:bugzilla_event} shows an example of such an \events{Event} and its corresponding \events{EventListener}.

\begin{lstlisting}[caption={\ac{RDF} representing an event generated by the creation of a bug.}, label={listing:bugzilla_event}]
    @prefix oslc: <http://open-services.net/ns/core#> .
    @prefix oslc_auto: <http://open-services.net/ns/auto#> .
    @prefix oslc_eca: <http://open-services.net/ns/eca#> .
    @prefix dcterms: <http://purl.org/dc/terms/> .
    @prefix xsd: <http://www.w3.org/2001/XMLSchema#> .
    @prefix example: <http://subclass.example.com#> .

    <http://bugzilla.example.com/ServiceProvider/1/Events/1> a oslc_eca:Event ,
            a example:BugCreationEvent ;
        oslc:serviceProvider <http://bugzilla.example.com/ServiceProvider/1> ;
        oslc_eca:generatedBy <http://bugzilla.example/EventListener/1> ;
        oslc_eca:causedBy <http://bugzilla.example.com/ServiceProvider/1/bugs/1> ;
        dcterms:created "2021-07-14T07:07:33+00:00"^^xsd:dateTime ;
        dcterms:title "Event generated by the creation of a bug"^^xsd:string .
\end{lstlisting}


The \ac*{OSLC} Bugzilla adapter also generates a changelog for the \ac*{TRS} server (Step 3).
It contains two \trs{Creation} resources pointing at the bug created and the event resource generated by the event listener.
The changelog is shown in Listing~\ref{listing:trs}.

\begin{lstlisting}[caption={\ac{RDF} representing the \ac*{TRS} logs that register the creation of the bug and its consequent event.}, label={listing:trs}]
    @prefix trs: <http://open-services.net/ns/core/trs#> .
    @prefix xsd: <http://www.w3.org/2001/XMLSchema#> .

    <http://bugzilla.example.com/trs/ChangeLog/1> a trs:ChangeLog ;
        trs:change [
            a trs:Creation ;
            trs:order "1"^^xsd:integer ;
            trs:changed <http://bugzilla.example.com/ServiceProvider/1/bugs/1> .
        ] ;
        trs:change [
            a trs:Creation ;
            trs:order "2"^^xsd:integer ;
            trs:changed <http://bugzilla.example.com/ServiceProvider/1/Events/1> .
        ] .
\end{lstlisting}

The event is then pushed to the \ac*{OSLC} Automation Server (Step 4), which sends it to the \ac*{TAS} in the corresponding format.
The \ac*{TAS} evaluates its automation rules and returns an action to the \ac*{OSLC} Automation Server.
The \ac*{OSLC} Automation Server stores a changelog in its \ac*{TRS} server, registering the creation of the corresponding AssertedCondition (Step 5).
Listing~\ref{listing:trs_auto} shows an \ac{RDF} representation of this changelog.

\begin{lstlisting}[caption={\ac{RDF} representing the \ac*{TRS} log message generated at the Automation Server.}, label={listing:trs_auto}]
    @prefix oslc_eca: <http://open-services.net/ns/eca#> .
    @prefix trs: <http://open-services.net/ns/core/trs#> .
    @prefix xsd: <http://www.w3.org/2001/XMLSchema#> .

    <http://auto.server/AssertedCondition/1> a oslc_eca:AssertedCondition ;
            oslc_eca:asserts <http://auto.server/ServiceProvider/1/Rule/1> ;
            oslc_eca:produces <http://auto.server/Action/1> ;
            oslc_eca:asserted "true"^^xsd:boolean .

    <http://auto.server/trs/ChangeLog/1> a trs:ChangeLog ;
        trs:change [
            a trs:Creation ;
            trs:order "1"^^xsd:interger ;
            trs:changed <http://auto.server/AssertedCondition/1> .
        ] .
\end{lstlisting}


Then, the \ac*{OSLC} Automation Server sends an action back to the Apache Kafka server (Step 6).
The \auto{status} and \auto{verdict} properties also indicate that it is still in process of being executed.
For this specific \textit{example:CreateIssueAction}, which is described by a ResourceShape not included for the sake of brevity, the \textit{example:originalChangeRequest} property links to the bug created previously in Bugzilla, which will be the source of information when GitHub's Action Dispatcher would execute the action.
Listing~\ref{listing:github_action} shows the Notation 3 code for this example \actions{Action}.

\begin{lstlisting}[caption={\ac{RDF} representing an action sent for execution to create a GitHub issue.}, label={listing:github_action}]
    @prefix oslc: <http://open-services.net/ns/core#> .
    @prefix oslc_eca: <http://open-services.net/ns/eca#> .
    @prefix oslc_auto: <http://open-services.net/ns/auto#> .
    @prefix oslc_cm: <http://open-services.net/ns/cm#> .
    @prefix dcterms: <http://purl.org/dc/terms/> .
    @prefix xsd: <http://www.w3.org/2001/XMLSchema#> .
    @prefix example: <http://subclass.example.com#> .

    <http://auto.server/Action/1> a oslc_eca:Action,
            example:CreateIssueAction ;
        oslc:serviceProvider <http://github.example.com/ServiceProvider/1> ;
        oslc_actions:executedBy <http://github.example.com/ActionDispatcher/1> ;
        oslc_auto:status oslc_auto:new ;
        oslc_auto:veredict oslc_auto:unavailable ;
        example:originalChangeRequest <http://bugzilla.example.com/ServiceProvider/1/bugs/1> ;
        dcterms:created "2021-07-14T07:07:33+00:00"^^xsd:dateTime ;
        dcterms:title "Action to be executed on GitHub"^^xsd:string .
        
    <http://auto.server/Action/1/versions/1>  dcterms:isVersionOf    
            <http://auto.server/Action/1> .
\end{lstlisting}



The action is delivered to the Action Dispatcher of the GitHub \ac*{OSLC} adapter (Step 7).
It then modifies the action's status and sets it to queued.
A \ac{TRS} changelog is generated representing this update.
It uses \ac{TRS} patch to show what properties changed and is displayed in Listing~\ref{listing:action_update}.

\begin{lstlisting}[caption={\ac{RDF} of a \ac{TRS} changelog showing the changes in an action status.}, label={listing:action_update}]
    @prefix xsd: <http://www.w3.org/2001/XMLSchema#>.
    @prefix dcterms: <http://purl.org/dc/terms/> .
    @prefix trs: <http://open-services.net/ns/core/trs#>.
    @prefix trspatch: <http://open-services.net/ns/core/trspatch#>.
    @prefix oslc_auto: <http://open-services.net/ns/auto#> .
    
    <http://github.example.com/trs/ChangeLog/1> a trs:ChangeLog ;
        trs:change [
            a trs:Creation;
            trs:changed <http://auto.server/Action/1/version/2>;
            trs:order "2"^^xsd:integer;
            trspatch:createdFrom <http://auto.server/Action/1/version/1>;
            trspatch:rdfPatch
                """
                D <http://auto.server/Action/1/versions/1>  dcterms:isVersionOf    
                        <http://auto.server/Action/1> .
                A <http://auto.server/Action/1/versions/2>  dcterms:isVersionOf          
                        <http://auto.server/Action/1> .
                D <http://auto.server/Action/1> oslc_auto:status oslc_auto:new.
                A <http://auto.server/Action/1> oslc_auto:status oslc_auto:queued.
                """ .
            ] .
\end{lstlisting}

The adapter then uses GitHub's API to create the issue described by the action parameters (Step 8).
Listing~\ref{listing:issue} showcases the newly created issue \ac*{RDF} representation.

\begin{lstlisting}[caption={\ac{RDF} representing the issues created in GitHub.}, label={listing:issue}]
    @prefix dcterms <http://purl.org/dc/terms/> .
    @prefix oslc_cm: <http://open-services.net/ns/cm#> .
    @prefix oslc: <http://open-services.net/ns/core#> .
    @prefix xsd: <http://www.w3.org/2001/XMLSchema#> .

    <http://github.example.com/ServiceProvider/1/issue/1> a oslc_cm:ChangeRequest ;
        oslc_cm:status "open"^^xsd:string ;
        oslc:serviceProvider <http://github.example.com/ServiceProvider/1 ;
        dctermscontributor "admin"^^xsd:string ;
        dcterms:created "2021-12-10T07:07:33+00:00"^^xsd:dateTime ;
        dctermsidentifier "1"^^xsd:integer ;
        dcterms:modified "2021-12-10T07:07:33+00:00"^^xsd:dateTime ;
        dcterms:title "Testing bug to issue automation"^^xsd:string .
\end{lstlisting}

Finally, the adapter sends the corresponding changelog to its \ac*{TRS} server (Step 9), registering the creation of the issue, as shown in Listing~\ref{listing:issue_trs}.
It also signals the update of the action to have its \actions{executedOn} property pointing to the appropriate resource, which was just created.

\begin{lstlisting}[caption={\ac*{RDF} representing a \ac*{TRS} log registering the correct execution of the action.}, label={listing:issue_trs}]
    @prefix trs: <http://open-services.net/ns/core/trs#> .
    @prefix oslc_auto: <http://open-services.net/ns/auto#> .
    @prefix xsd: <http://www.w3.org/2001/XMLSchema#> .

    <http://github.example.com/trs/ChangeLog/2> a trs:ChangeLog ;
        trs:change [
            a trs:Creation ;
            trs:order "1"^^xsd:integer ;
            trs:changed <http://github.example.com/ServiceProvider/1/issue/1> .
        ] ;
        trs:change [
            a trs:Modification ;
            trs:order "2"^^xsd:integer ;
            trs:changed <http://auto.server/Action/1> ;
            trspatch:createdFrom <http://auto.server/Action/1/version/2>;
            trspatch:rdfPatch
                """
                D <http://auto.server/Action/1/versions/2>  dcterms:isVersionOf    
                        <http://auto.server/Action/1> .
                A <http://auto.server/Action/1/versions/3>  dcterms:isVersionOf          
                        <http://auto.server/Action/1> .
                D <http://auto.server/Action/1> oslc_auto:status oslc_auto:queued.
                A <http://auto.server/Action/1> oslc_auto:status oslc_auto:completed.
                D <http://auto.server/Action/1>      oslc_auto:veredict oslc_auto:unavailable ;
                A <http://auto.server/Action/1>      oslc_auto:veredict oslc_auto:passed ;
                A <http://auto.server/Action/1> oslc_eca:executedOn
                        <http://github.example.com/ServiceProvider/1/issue/1>.
                """ .
            ] .
        ] .
\end{lstlisting}

To summarize, the worked example presents a practical implementation of the architecture based on the proposed \ac*{OSLC} extension to support events and actions. It meets the requirements to address the common issues faced when adopting DevOps. Because both services are integrated using \ac*{OSLC} interfaces, requirement RQ1 is fulfilled, making integrating new services more straightforward and efficient. Automation is separated into its own service accessible via \ac*{OSLC}, satisfying requirement RQ2. Traceability is managed with the \ac*{TRS} protocol, a standard within \ac*{OSLC}, and meets requirement RQ3. Finally, requirement RQ4 is fulfilled because the components are loosely coupled, and a distributed service handles messaging, allowing the architecture to scale horizontally.

\section{Conclusions and future work}\label{sec:conclusions_and_future_work}

This paper presents an extension model for the \ac*{OSLC} standard to support \ac*{ECA}-based automation. More specifically, it provides the concepts to make \ac*{ECA} automation possible in an interoperable environment. The end goal of this proposal is to use \ac*{OSLC} to improve the adoption of DevOps by facilitating integration and automation between tools from different vendors or domains.

The proposed extension model has been validated by studying the main issues in the DevOps field, and the architectural characteristics deemed beneficial by practitioners. Next, a set of requirements has been established based on the said issues and used for validation. Then, a prototype architecture was designed to assess the requirements of the motivational environment. Finally, a worked example has been conducted that involves two major software development tools, GitHub and Bugzilla, to detail the application of the proposed model in a real-world workflow.



The contribution is expected to be helpful in automated environments where interoperability is a priority. Moreover, the semantic model could potentially contribute to the open project of the \ac*{OSLC} standard.

For future work, a methodology for adopting \ac*{OSLC} and \ac*{ECA} automation for DevOps professionals will be developed. This will make the architecture easier to translate into a real-world scenario. Also, other kinds of service (beyond the change management domain) will be integrated into the architecture to test more characteristics relevant to DevOps and to provide a more complex and interesting worked example. For example, tools used to manage infrastructures, such as Docker~\citep{Docker} or Kubernetes~\citep{Kubernetes}, are very popular in DevOps environments. They can help manage large architectures efficiently and accelerate the deployment of services.
Standardizing their interfaces with \ac*{OSLC} could help integrate them with even more tools and enable event-based automation.

Another aspect of \ac*{ECA} that will be explored in the future is automation rules. Semantic web technologies have also been proposed in this domain. A bridge between one \ac*{ECA} ontology, such as \ac*{EWE}~\citep{coronadoModellingRulesAutomating2015}, and the proposed model would allow for more powerful automation features.

Furthermore, other issues concerning DevOps are explored in the systematic review used to establish the requirements of the prototype architecture~\citep{bolscherDesigningSoftwareArchitecture2019}. These challenges could be explored in future work, such as testing and monolithic databases~\citep{bolscherDesigningSoftwareArchitecture2019}. In addition, the model can still be extended to support new features and domains. For example, automation infrastructures often provide a login system for users that has not been implemented in this use case. The vocabulary can be extended further with new concepts to cover all these aspects.

\bibliography{references}
\bibliographystyle{plainnat}

\end{document}